# Albe 1998 – La grande Motte 2009 : quelles avancées en 10 ans ?

Isabelle Grillo

*Institut Laue Langevin, DS-LSS, 6 rue Jules Horowitz, B.P. 156, 38042 Grenoble Cedex 9*

**Abstract.** The importance of neutron scattering techniques for the characterization of samples in soft condensed matter has been demonstrated all along the present book. The fine understanding of the physical properties is closely linked to progress in the field of instrumentation. This chapter describes the advances over the last decade in technical domains, such as neutron detection, electronics and sample environment. The news software for data reduction and analysis are also discussed before to conclude with the ILL and LLB projects for new instruments.

**Résumé.** Tout au long de ce livre, nous avons vu l'importance des techniques de diffusion de neutrons pour la caractérisation des échantillons en matière molle. La compréhension toujours plus détaillée de la physique des échantillons est intimement liée aux progrès instrumentaux. Ce chapitre présente les avancées et les développements réalisés pendant ces dix dernières années dans les domaines techniques, comme ceux de la détection, de l'électronique et des environnements échantillons. Les nouveaux moyens de traitement et d'analyse des données sont ensuite présentés. Nous terminons par les projets instrumentaux de l'ILL et du LLB.

Depuis 30 ans, les techniques de diffusion, en particulier la DNPA et plus récemment la réflectométrie ont indiscutablement contribué aux avancées des recherches dans le domaine de la matière molle. Toujours extrêmement demandées (il y a en moyenne un facteur 3 à 4 entre le nombre de jours souhaité et celui attribué dans les comités de sélection), ces expériences peuvent apparaître comme « routinières ». Pourtant, elles sont de plus en plus pointues et permettent des représentations et des compréhensions des systèmes expérimentaux de plus en plus précises. On observe ainsi une évolution parallèle de la complexité des techniques mises en œuvre et de celle des systèmes étudiés. Nous essayerons de mettre en évidence dans ce chapitre les progrès les plus marquants de ces 10 dernières années.

## 1. Des progrès majeurs dans le domaine de la détection et de l'électronique

Malgré le flux relativement faible des faisceaux de neutrons comparés à ceux des rayons X produits par un synchrotron, il est parfois nécessaire pour des échantillons fortement diffusant d'atténuer le flux pour ne pas saturer le détecteur. L'électronique d'acquisition et le temps de collecte des charges sont propres à chaque détecteur et fixent un temps mort, qui est l'intervalle de temps après la détection d'un neutron pendant lequel un autre neutron ne peut être détecté. Le taux maximal de comptage dépend ensuite de ce temps mort. La technologie des détecteurs d'il y a 10 ans présentait un temps mort de l'ordre de 1 µs et permettait un comptage jusqu'à 80-100 kHz avec une perte de 10%, qu'il était possible de corriger grâce à des méthodes mathématiques [1]. Au-delà, les mesures n'étaient plus possibles. De grands progrès ont été d'abord réalisés dans l'électronique de détection[*] réduisant le temps mort d'un facteur 3 à 4. De nouveaux types de détecteurs ont ensuite été développés. Le détecteur installé sur D22 (ILL, France) en 2004 est composé de 128 tubes alignés verticalement. Chaque tube est un compteur indépendant capable d'atteindre 80 kHz avec une perte de 10%. Pour une diffusion isotrope (sans « spot » de type pic de Bragg), il est maintenant possible de mesurer jusqu'à 2 MHz, sans correction de temps mort. En dix ans, un facteur 200 en taux de comptage a donc été gagné.

Une utilisation optimale du flux des instruments est maintenant possible, réduisant les temps d'acquisition ou permettant pour un même temps de mesure, un meilleur rapport signal sur bruit. Ceci est particulièrement vrai aux grands angles, lorsque la diffusion totale est importante, mais essentiellement due au bruit de fond incohérent. L'analyse fine des petits objets, comme les micelles, les épaisseurs de membranes ou les protéines est donc fortement améliorée.

Une autre limitation des détecteurs actuels pour la diffraction de monocristaux, mais aussi pour la réflectométrie et la DNPA est la résolution spatiale. Avec le projet européen MILAND (Millimetre Resolution Large Area Neutron Detector) arrive une nouvelle génération de détecteurs avec une surface de détection de 32x32 $cm^2$, une résolution spatiale de 1 $mm^2$ et un taux de comptage jusqu'à 1 MHz pour un temps mort de 10% [2]. Déjà testé avec succès en 2008, MILAND sera installé sur l'instrument D16 à l'ILL fin 2010.

## 2. Imaginer de nouveaux environnements échantillons

La nature en elle-même est muette, il faut donc imaginer des moyens pour qu'elle nous parle. Ainsi, dans les 10 dernières années, on a observé le développement de nouveaux environnements expérimentaux, adaptés à partir de techniques déjà existantes ou spécialement conçus pour la DNPA. Ce paragraphe présente quelques récents développements maintenant couramment utilisés. Dans beaucoup de ces exemples, les utilisateurs des grands instruments sont fortement impliqués dans la conception.

### 2.1 Mesures sous cisaillement

Par leur nature propre, les échantillons en matière molle sont particulièrement sensibles aux forces de cisaillement et d'écoulement. Comprendre et prédire les réarrangements des structures à l'échelle moléculaire et macroscopique représente un enjeu majeur aussi bien fondamental qu'industriel dans le domaine des polymères, des colloïdes et des surfactants. Par exemple, les objets en plastique sont obtenus par moulage ou par extrusion, à hautes températures et pression. Ou encore, l'efficacité des lessives ou shampoings est liée à un fort cisaillement du liquide ou du gel, qui produit une mousse ou une émulsion. La viscosité, l'élasticité, les propriétés rhéologiques et mécaniques dépendent non seulement de la nature des molécules mais aussi des structures formées à plus grandes échelles. Un meilleur contrôle de ces structures pourrait permettre d'améliorer les propriétés finales des matériaux. Les premières expériences combinant rhéomètre (type Couette) et DNPA remontent à plus de 25 ans [3]. La cellule est composée d'une double enveloppe avec un piston intérieur fixe (stator) et d'un cylindre rotatif extérieur (rotor). L'échantillon remplit l'espace intermédiaire. Ce dispositif permet de mesurer la diffusion d'une solution soumise à un gradient de cisaillement presque constant. Le faisceau de neutron peut passer soit radialement, dans la direction du gradient, soit de face, dans la direction perpendiculaire au gradient, ce qui permet de sonder les structures dans deux plans différents pour une caractérisation complète.

---

[*] http://www.ill.eu/instruments-support/instruments-groups/instruments/d22/more/fast-detector/electronics/

Au début des années 2000, Bent *et col* ont spécialement fabriqué un extrudeur adapté à la géométrie de D22 à l'ILL pour l'étude des fondus de polymère, soumis à des contraintes de contraction puis d'expansion. La géométrie de la cellule est représentée sur la figure 1. La solution de polymère fondu circule à haute température et haute pression grâce à une pompe à engrenage. A son entrée, la cellule d'observation mesure 10 mm de large ; elle se réduit brusquement à 2,5 mm, provoquant une contraction de la solution sur 20 mm de longueur, puis elle retrouve en sortie une largeur de 10 mm [4]. Le débit peut atteindre 6 $cm^3.s^{-1}$, pour des températures allant jusqu'à 230°C, des pressions de 10 MPa (100 bar) et des viscosités de 65 kPa.s. Un diaphragme de 1 mm de large par 5 mm de hauteur définit la taille du faisceau. La cellule est translatée verticalement pour regarder la diffusion en différentes positions. Ce dispositif, combiné à une deutériation spécifique du polymère permet d'étudier la conformation des chaines en fonction des différents paramètres cités plus haut et de la position dans la cellule. Les différentes études se sont intéressées à l'influence de la masse molaire et de la polydispersité ainsi que les ramifications des chaines [5].

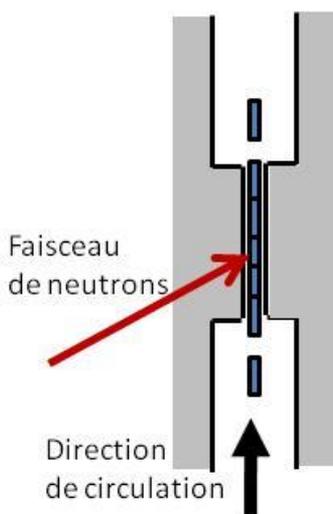

Figure 1 : représentation schématique de la cellule à circulation développée par Bent *et col* d'après [4]. Un petit diaphragme de 1x5 $mm^2$ placé devant l'échantillon permet de mesurer la diffusion des chaines à différents endroits de la cellule.

Plus récemment, Rennie *et col* ont mis au point une cellule à recirculation, ayant une géométrie en forme de croix, pour étudier l'alignement de colloïdes inorganiques anisotropes (plaquettes), soumis à un écoulement élongationnel (figure 2). Un petit diaphragme circulaire de 1 mm de diamètre délimite la taille du faisceau. La cellule est translatée verticalement et horizontalement pour une cartographie sous écoulement des structures dans la solution [6].

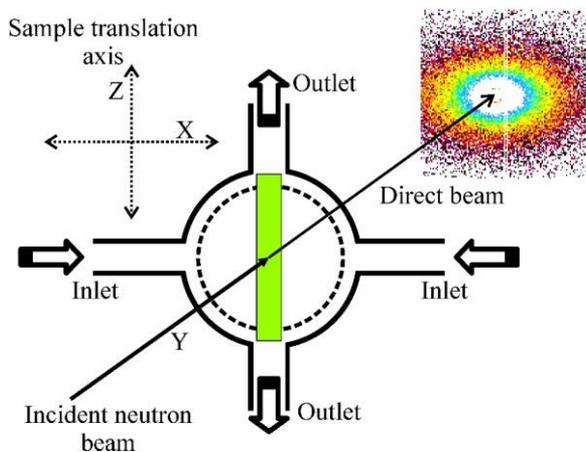

Figure 2 : Représentation schématique de la cellule pour l'étude d'un écoulement élongationnel d'après [6]. La cellule est translatée selon les directions X et Z pour obtenir une cartographie des structures. La région surlignée en vert délimite la zone où le maximum d'alignement des particules est observé. La zone délimitée par des pointillés représente la zone d'observation possible.

2.2   Mesures résolues en temps

Les instruments de l'ILL, avec un haut flux, jusqu'à $10^8$ n/s/cm$^2$ ont ouvert un domaine de recherches avec les mesures de cinétiques rapides, afin de suivre les transformations en temps réel d'un échantillon après une perturbation (dilution, mélange, saut de pH ou de température, modification de la force ionique, application d'un champ magnétique,…). La connaissance des toutes premières étapes de formation d'un échantillon et des possibles états intermédiaires peut être cruciale pour comprendre et agir sur l'état d'équilibre.

Les travaux pionniers de S. Egelhaaf ont suivi pendant plusieurs heures la transition entre micelles et vésicules induite par la dilution manuelle d'un mélange de lécithine et de sels biliaires [7]. Pour des temps beaucoup plus courts, de l'ordre de la seconde, il est capital de synchroniser la détection des neutrons diffusés avec l'environnement expérimental afin d'assurer la reproductibilité de l'expérience. Couplé à la diffusion de la lumière, à des spectroscopies IR ou UV ou à la RMN, un appareil à flux stoppé « stopped-flow » est un des équipements les plus utilisés pour suivre des cinétiques rapides. Il permet en un temps très court, de quelques ms, de mélanger plusieurs solutions et de synchroniser le début de la réaction avec la mesure. Cette technique est disponible depuis 2001 sur l'instrument D22 à l'ILL. Le temps minimum « raisonnable » par acquisition est de 50 à 100 ms. Dans la plupart des cas, l'expérience est répétée 5 à 10 fois et les spectres sont additionnés pour obtenir une statistique suffisante. Une description précise de la technique, ainsi qu'une revue des systèmes ainsi étudiés sont à lire dans la référence [8]

Contrairement au « flux-stoppé » où, comme le nom l'indique l'échantillon est stoppé et évolue dans la cellule d'observation, un autre type d'expérience en temps réel utilise une cellule à recirculation couplée à une pompe péristaltique. La solution passe donc à un débit défini par l'expérimentateur à travers la cellule d'observation et des mesures sont régulièrement faites. Ce principe a été utilisé par Imperor *et col* pour suivre les premiers stades de formation de matériaux mésoporeux [9]. En utilisation une deuteriation spécifique du solvant, les expériences de DNPA ont ainsi pu mettre en évidence les transformations géométriques de la matrice organique induites par l'absorption et l'association des espèces inorganiques.

## 2.3   Mesures sous pression osmotique contrôlée

Pour les échantillons de matière molle, un paramètre thermodynamique important est le taux d'hydratation, qui détermine la structure et la dynamique des objets. Communément, la pression de vapeur saturante est imposée par une solution saline saturée à une température donnée. Au début des années 2000, un nouveau type de chambre à humidité à été conçu sur l'instrument D16 (ILL). Elle comporte deux compartiments isolés thermiquement et maintenus à des températures différentes. Celui du bas contient un réservoir avec de l'eau ($H_2O$ ou $D_2O$), celui du haut l'échantillon. Un petit trou permet l'échange de vapeur entre les deux chambres. La différence de température entre les deux compartiments permet d'imposer la pression osmotique dans l'échantillon [10]. La difficulté de ce montage vient du contrôle précis de la température dans tout le volume des compartiments pour éviter les gradients de température. Cette chambre a été particulièrement utilisée pour des études du gonflement, de l'empilement et des fluctuations des membranes constituées de lipides ou glycolipides. Les plus récentes peuvent être lues dans les références suivantes [11,12,13].

Une autre cellule originale a été construite pour étudier les propriétés mécaniques et les interactions inter-membranaires de lipopolysaccharides, constituant la membrane externe de certaines bactéries [14]. La cellule liquide consiste en deux substrats de silicium ; sur l'un d'eux sont déposées les membranes biologiques, qui forment un empilement lamellaire. Les substrats sont séparés par des espaceurs en verre et les forces capillaires confinent une mince couche d'eau ou de tampon entre les deux substrats. Le tout est maintenu dans la chambre à humidité, pour contrôler la température et pour éviter l'évaporation. Un tel dispositif permet de travailler en excès d'eau ou de tampon et donc à pression osmotique nulle.

## 2.4   Cellule haute pression

Un autre paramètre important, mais peut-être moins souvent étudié pour des solutions « matière molle » est la pression. Durant ces 10 dernières années, plusieurs équipes ont développé leur cellule. Celle du LLB (Saclay, France), adaptée à la fois aux petits angles et la diffusion inélastique permet d'atteindre 7000 bar [15]. Au HMI (Berlin, Allemagne), la cellule pression permet d'atteindre 5000 bar pour un domaine de température de 5 à 80°C. Des sauts de pressions de 1000 bar en 1 seconde sont également possibles. Ses fenêtres en saphir permettent une mesure simultanée en diffusion de lumière [16]. Le montage expérimental développé à l'ILL (Grenoble, France) a été construit afin de réaliser des sauts de pression dans un temps le plus court possible de l'ordre de 100 ms pour suivre les cinétiques liées aux variations de pression [17]. Toutes ces cellules ont principalement pour application l'étude de molécules biologiques, comme les protéines. Pepy *et col.* ont également conçu une cellule atteignant 1500 bar et jusqu'à des températures de 423 K pour l'étude de cristaux liquides de polymères [18].

En matière molle l'eau reste le principal solvant utilisé. Néanmoins dans la recherche de nouveaux solvants, le $CO_2$ présente plusieurs avantages comparés aux solvants organiques. Il est plus respectueux de l'environnement, recyclable, ni toxique, ni inflammable et relativement bon marché. Ses propriétés de solvatation peuvent être modulées en fonction de la pression appliquée. Le $CO_2$ devient super critique ($CO_2sc$) à des températures proches de l'ambiante, au dessus de 31.1°C et une pression supérieure à 73.8 bar. Un dispositif constitué d'une cellule à volume variable équipée d'une fenêtre en saphir couplée avec une cellule haute pression en Niobium permettant de travailler sous température et pression contrôlée avec du $CO_2sc$, a récemment été mis au point [19] et utilisé pour étudier l'auto-organisation de copolymères fluorés. La cellule à volume variable permet également des scans rapides en pression. De nombreuses applications et études de DNPA s'ouvrent ainsi. En effet, le $CO_2$ étant considéré comme un solvant alternatif plus écologique, la synthèse et caractérisation de nouvelles molécules CO2-philes est un domaine extrêmement actif [20].

## 2.5   Environnement pour application spécifique

Enfin, comme dernier exemple de nouvel environnement échantillon, on peut trouver dans ce livre la description d'un prototype de pile à combustible spécialement conçu pour pouvoir être mis dans un faisceau de neutrons afin de caractériser les profils de concentration d'eau à travers une membrane au cours de son fonctionnement (chapitre de Sandrine Lyonnard et ref 21)

## 2.6   Environnement pour la réflectivité

L'importance de l'interface air/eau en matière molle ou biologie a récemment motivé la construction de nombreux réflectomètres horizontaux (FIGARO à l'ILL à Grenoble, REFSANS au FRM2 à Munich en Allemagne, SNS-LR à Oak Ridge aux USA) complétant ceux déjà existant au LLB en France (EROS) ou à ISIS en Grande-Bretagne (CRISP et SURF). Tous ces appareils proposent désormais des mesures in-situ sur des cuves de Langmuir, qui sont incontournables pour la physique 2-D à l'interface air/eau.

## 3. Des programmes de pilotage et de traitements intelligents

L'amélioration des performances des instruments de diffusion est étroitement liée au développement de nouveaux programmes de contrôle des instruments, de réduction et d'analyse des données.
Les nouveaux programmes de pilotage « intelligents » n'ont plus seulement pour but de mettre l'instrument dans une configuration souhaitée par l'expérimentateur (distance échantillon-détecteur, collimation, longueur d'onde, position de l'échantillon) mais doivent pouvoir piloter et contrôler les acquisitions. Une question essentielle est l'optimisation du temps de faisceau et de la qualité des données enregistrées. Ceci est particulièrement important dans le cas d'une série d'échantillons, diffusants très différemment. Actuellement, il faut manuellement mesurer rapidement (de la seconde à la minute) chaque échantillon afin d'estimer le taux de comptage et de calculer le temps nécessaire. Dans un programme évolué, on pourrait faire correspondre à chaque échantillon sa cuve vide, son solvant et le bruit de fond pour qu'une analyse rapide calcule la statistique moyenne ou dans une région d'intérêt et stoppe l'acquisition lorsque cette dernière, préalablement définie par l'expérimentateur est atteinte. Ces développements sont particulièrement importants dans la région des grands angles et de Porod, lorsque le signal incohérent domine la diffusion. Il arrive fréquemment qu'après normalisation et soustraction, et pourtant avec plusieurs millions de coups sur le détecteur, le signal soit très bruité et l'analyse difficile.
Les mesures de DNPA génèrent généralement un très grand nombre de données. Dans des modes cinétiques on enregistre plusieurs milliers de spectres en deux ou trois jours de temps de faisceau. La réduction des données isotropes, c'est-à-dire la transformation d'une image bidimensionnelle en une courbe où l'intensité en $cm^{-1}$ est tracée en fonction du vecteur d'onde q est décrite étape par étape dans le cours de D. Lairez de ce livre. C'est une succession simple mais rigoureuse d'opérations mathématiques qui deviennent vite fastidieuses si elles doivent être faites manuellement. Les différents instituts ont développé de nouveaux programmes très automatisés, capables de faire un tri des échantillons par configuration, de déterminer les centres des faisceaux et les masques, de tenir compte des corrections géométriques nécessaires aux grands angles et finalement de normaliser les données. On peut citer BerSANS pour le HMI [22], Grasp [23] et Lamp [24] pour l'ILL et PAsiNET [25] pour le LLB développés dans ces dernières années. Ils permettent un traitement en ligne des données et offrent ainsi la possibilité de vérifier la qualité des échantillons et la préparation ciblée de nouveaux échantillons en fonction des résultats obtenus pour une optimisation du temps de faisceau. Pour une meilleure collaboration des scientifiques entre les communautés neutrons, rayon-X et muons autour du monde, un format standard et international, appelé NeXus se met progressivement en place [26]. Il permettra de visualiser et d'analyser les données en utilisant de façon indifférenciée les différents programmes des instituts.

L'analyse des données pour en extraire les paramètres physiques de l'échantillon nécessite des modèles mathématiques, beaucoup sont disponibles depuis plusieurs dizaines d'années pour les formes et les structures les plus classiques. On peut distinguer deux approches dans l'analyse, la première est dite « indépendente du modèle », la seconde est appelée « modélisation directe ».
Une méthode classique de la première catégorie consiste en une transformée de Fourier inverse de la courbe expérimentale, qui donne accès à la fonction de distribution de densité de paires p(r) (pair density distribution function, PDDF). Cette opération est faite par la méthode de transformée de Fourier indirecte (IFT), introduite en 1977 par O. Glatter [27]. Dans le cas de systèmes dilués aléatoirement orientés, la fonction p(r) donne des informations sur la taille et la forme des objets diffusants. Dans le cas d'un objet centro-symétrique, le profil de densité de longueur de diffusion peut être obtenu. Pour les systèmes en interaction, la transformée de Fourier indirecte généralisée (GIFT) introduite en 1997 permet séparer les contributions intra et inter objets et donc de déterminer à la fois la forme et la structure [28].
Des programmes de D. Svergun basées sur des techniques de recuit simulé sont développés pour des échantillons en solution dont la structure atomique est connue mais pas l'organisation à grande échelle [29,30]. Le système est décrit par une large assemblée compacte de billes, qui appartiennent soit aux particules, soit au solvant [31]. Des contraintes imposées par la physique du système sont introduites afin de réduire le nombre de degrés de liberté. Un recuit simulé est utilisé pour rechercher le modèle compact qui simule le mieux les données expérimentales. Ces programmes régulièrement mis à jour sont donc largement utilisés en biologie pour la modélisation des structures ternaires et quaternaires de protéines, ou encore la caractérisation de complexes macromoléculaires [32].
Les techniques de simulation Monte Carlo et de dynamique moléculaires sont également très utilisées, particulièrement dans les cas où il n'existe pas de solution analytique rigoureuse, comme pour les polymères [33,34] ou des systèmes mixtes complexes [35,36]. De façon complémentaire, la technique de Monte Carlo reverse (RMC) a été très largement développée par McGreevy et Pusztai [37,38]

Lorsque les expressions analytiques existent, on parle alors de modélisation directe. De nombreuses références décrivent les modèles les plus couramment utilisés, il serait ici impossible d'en faire une liste exhaustive. Néanmoins, on peut se référer à la documentation du programme d'analyse de DNPA produit par le NIST [39] où les modèles sont mathématiquement décrits avec les références des papiers fondateurs.
Depuis quelques années, de nombreux logiciels incluant facteur de forme et de structure, polydispersité et résolution instrumentale, proposant une interface graphique et de nombreuses fonctionnalités sont disponibles. Testés et comparés, ils assurent une bonne fiabilité et évitent à chaque chercheur de réécrire son propre code. Ils sont particulièrement importants pour les nouveaux utilisateurs des techniques de diffusion. On peut citer pour les petits angles : Fish (ISIS, Grande-Bretagne), SASfit (PSI, Suisse), DANSE (University of Tennessee, Etats-Unis, [40]) lui-même basé sur les modèles développés au NIST [39].

La complexité des systèmes étudiés maintenant, associant plusieurs colloïdes ou différentes phases nécessitent bien sûr l'élaboration de nouveaux modèles et équations. A l'initiative de J. Kohlbrecher (PSI, Suisse), il est fortement question de créer une base de données afin de faire bénéficier toute la communauté des modèles publiés, pour une optimisation de l'analyse des données.
Comme pour la réduction de données, une étape supplémentaire devra être franchie pour l'automatisation de l'analyse, pour les expériences qui génèrent des milliers de fichiers, comme les cinétiques, et les mesures stroboscopiques du futur instrument « Tisane ». Une caractéristique de ces données est qu'elles n'évoluent pas de façon arbitraire. Par exemple on observe des objets qui grossissent en gardant la même forme, ou des transitions de phases, le volume de l'une augmentant aux dépends de l'autre. Les futurs logiciels devront être capables de lire une série de données, puis d'appliquer un modèle en variant régulièrement un ou deux paramètres caractéristiques. Sans ces développements, il est à craindre que moins d'un pour-cent des données issues des ces nouvelles techniques soit exploité !

## 4. De nouveaux instruments pour la matière molle

Dans ce paragraphe, nous nous intéresserons plus particulièrement aux instruments des centres de diffusion de neutroniques français (ILL et LLB). Néanmoins, pour les lecteurs intéressés une liste exhaustive des instruments « petits angles » dans le monde et des scientifiques impliqués peut être trouvée sur le lien suivant :
http://www.ill.eu/instruments-support/instruments-groups/groups/lss/more/world-directory-of-sans-instruments/

En 2001, l'ILL a lancé la première phase du programme « millénium » de modernisation. Le spectromètre D22 en a bénéficié avec le développement et l'installation d'un détecteur rapide, comme nous l'avons vu plus haut dans ce chapitre. Sur l'instrument D11, la collimation a entièrement été refaite. Juste après le sélecteur un guide trapézoïdal de 6,5 m de longueur permet d'augmenter la section du faisceau de 30x50 mm$^2$ à 45x50 mm$^2$. Symétriquement les trois dernières sections du guide focalisent le faisceau sur une surface de 31,5x35 mm$^2$. Deux nouvelles sections de guide ont été ajoutées à des distances de 28,5 et 34 m, augmentant ainsi le flux pour les mesures aux très petits angles et permettant à 34 m et λ=6 Å un recouvrement avec le domaine de la lumière. Le tube détecteur a été remplacé pour pouvoir accueillir un détecteur à gaz plus large, d'un mètre carré de surface qui avec sa nouvelle électronique peut atteindre des taux de comptage de 300 kHz.

Le réflectomètre horizontal FIGARO (Fluid Interface Grazing Angles ReflectOmeter) a accueilli ses premiers utilisateurs en Avril 2009. Il est plus particulièrement destiné aux échantillons matière molle et biologie et sa géométrie horizontale permet des mesures à l'interface air/liquide. Ses caractéristiques sont un flux élevé grâce à des guides de neutrons super-miroirs et une grande flexibilité dans le choix de la résolution et des configurations.

Pendant la seconde phase de modernisation (2008-2014) le projet majeur en matière molle sera la construction du troisième instrument de diffusion de neutrons aux petits angles D33. Il fonctionnera soit dans un mode monochromatique avec un sélecteur de vitesse, soit en mode temps de vol, avec un système de double choppers qui permettra d'étendre le domaine de vecteur d'onde mesuré en une configuration et une grande flexibilité sur la résolution en longueur d'onde. Les mesures simultanées sur deux détecteurs multitubes permettront d'obtenir en une seule mesure un domaine dynamique $q_{max}/q_{min}$ de 20 en mode monochromatique et jusqu'à 1000 en mode TOF (Temps-de-Vol) [41]. Cet instrument sera également dédié à l'étude fine de structures magnétiques grâce à un faisceau polarisé couplé à un analyseur de spin à $^3$He ( $^3$He spin analysis ).

Au LLB, le nouveau spectromètre TPA (Très Petits Angles ou VSANS en anglais) présente des caractéristiques uniques et originales. Pour atteindre une haute résolution combinée avec la plus petite valeur de q possible, et une distance échantillon détecteur relativement courte (6 m) et pour des raisons d'espace disponible dans le hall des guides, un multi-détecteur avec des pixels de très petite taille était nécessaire [42]. Ces contraintes ont orienté le choix vers un détecteur « Image plate » développé initialement pour les rayons X et équipé d'un convertisseur neutrons. Il comporte 2300x2300 pixels de 0,15 x 0,15 mm$^2$. Le principal inconvénient est sa sensibilité au rayonnement γ. Il n'est alors plus possible d'utiliser un sélecteur de vitesse, qui à cause du Gadolinum ou du Bore utilisé comme absorbeur produit une forte radiation γ. Un monochromateur composé de deux super-miroirs parallèles a donc été conçu. Une double réflexion permet la monochromatisation du faisceau qui sort latéralement déplacé par rapport au faisceau incident, ce qui permet de bloquer tout rayonnement parasite venant de guides. La longueur d'onde varie de 5 à 20 Å en fonction de l'angle de rotation des super-miroirs. La résolution est fixée par les caractéristiques des miroirs. Sa collimation est composée de masques percés de petits trous (multi-diaphragmes) de 1 mm de diamètre, produisant ainsi de très petits multi-faisceaux convergents en un point sur le détecteur. Les masques intermédiaires sont positionnés de manière à éviter les interférences entre les différents canaux virtuels. Pour les premières expériences réalisées [43], 7 masques percés de 51 trous, de 1,7 mm de diamètre sur le premier masque et de 1 mm sur le dernier, organisés sur un réseau hexagonal ont été installés. Il est ainsi possible de descendre à des vecteurs d'onde de $2 \times 10^{-4}$ Å$^{-1}$ avec une très grande résolution en q. Le q max est de l'ordre de $10^{-2}$ Å$^{-1}$, assurant ainsi un bon recouvrement avec les instruments classiques de DNPA.

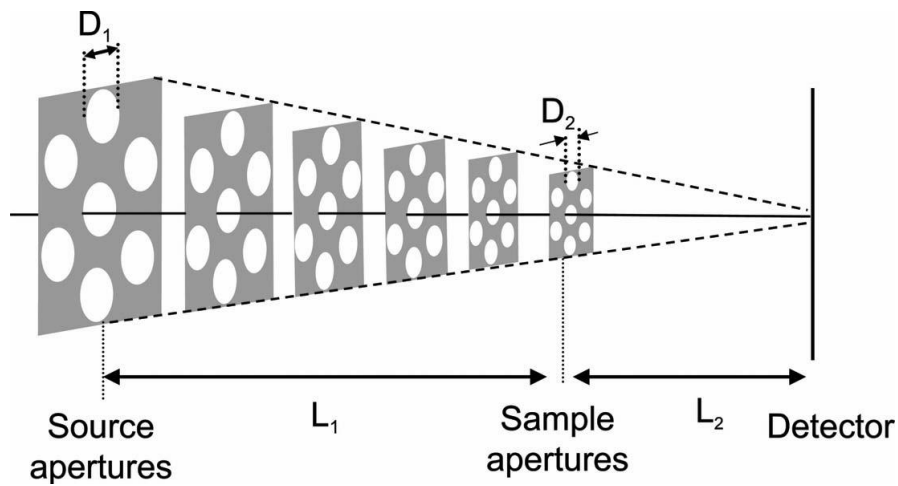

Figure 3 : Représentation schématique de la collimation de TPA composée de masques percés et focalisant sur le détecteur d'après [43]. Les masques intermédiaires sont placés de manière à supprimer les interférences entre les différents canaux.

Un second projet important du LLB pour la matière molle est la reconstruction de PAXE qui deviendra PA20, avec une collimation de 20 m de longueur et 20 m de distance entre l'échantillon et le détecteur ainsi qu'une option neutrons polarisés. Un nouveau détecteur multi-tubes de 64 cm$^2$ x 64 cm$^2$ de surface (selon la technologie développée à l'ILL) sera installé, et potentiellement un second permettra de mesurer simultanément les grands angles et les petits angles. La géométrie et le « design » de PA20 et D33 seront donc très similaires. Des systèmes de focalisation comme des lentilles ou des miroirs convergents sont envisagés pour augmenter le flux.

En effet, une question fondamentale et récurrente en diffusion de neutrons est comment augmenter le flux tout en gardant une résolution correcte. Nous avons vu la solution proposée avec les multi-diaphragmes pour TPA au LLB. Une autre consiste en une série de lentilles biconcaves convergentes en MgF$_2$ placées juste avant l'échantillon et permet d'obtenir un gain en flux d'un ordre de grandeur [44,45]. Des lentilles magnétiques ont également été mises en place sur SANS-J au Japon [46]. Ces dispositifs permettent un gain de flux, une meilleure résolution et un plus petit q$_{min}$ mais aux dépends de la flexibilité de l'instrument qui n'est alors optimisé que pour une longueur d'onde ou une distance-échantillon détecteur [47].

Par ailleurs, les performances du réflectomètre EROS [ref : http://www-llb.cea.fr/fr-en/pdf/eros-llb.pdf] du LLB ont été fortement améliorées par une refonte progressive et totale de l'appareil depuis 10 ans. C'est le réflectomètre servant pour toutes les applications de matière molle au laboratoire. Il est à géométrie horizontale, permettant donc les études à l'interface air/liquide, et fonctionne en temps de vol. Dans un premier temps, le chopper original mono-disque a été remplacé par un chopper multidisques permettant de varier la résolution (voir le cours de G. Fragneto du présent ouvrage pour le principe) et permettant de gagner un facteur d'efficacité de ~5 sur les taux de comptage par rapport à l'ancien chopper détecteur dans la résolution la plus dégradée. Le collimateur a ensuite été remplacé par un nouveau collimateur court ayant permis de gagner un facteur d'efficacité de ~10 sur les taux de comptage. Grâce à ces améliorations, il a été possible récemment de faire des mesures jusqu'à R~$10^{-6}$ sur des échantillons de 15 mm$^2$ ! Enfin le monocompteur utilisé actuellement va être remplacé par un multicompteur 2D testé actuellement qui permettra les mesures hors-spéculaires et qui sera proposé aux utilisateurs de façon routinière au printemps 2011. Par ailleurs, différents environnements échantillons ont été testés avec succès ces dernières années sur l'appareil pour des mesures in-situ (rhéomètre, chambre à humidité contrôlée, cuve de Langmuir) offrant aux utilisateurs une large gamme de possibilités de géométrie des mesures.

## 5. Pour conclure…

En dix ans, les progrès techniques et informatiques ont été considérables, relançant les recherches dans les domaines déjà bien établis et ouvrant de nouveaux champs d'application. Dans tous les instituts les instruments de diffusion de neutrons aux petits angles font partie de ceux les plus demandés. Mais malgré tout ces développements qui apportent confort pendant l'expérience, simplicité et gain de temps lors de la réduction du traitement et de l'analyse des données, la présence de l'expérimentateur et son expertise resteront toujours irremplaçables !

**Références**

[1] Knoll G.F. (1989) Radiation Detection and Measurement, Chap. 7, 2nd ed.Wiley, New York;P. Lindner, , ILL Technical report, (1998) ILL98/LI 12 T
[2] Buffet J.C. et al. Nuclear Instruments and Methods in Physics Research A (2005) **554,** 392–405
[3] Lindner P. and Oberthür R.C., Rev. Phys. Appl. (1984) **19,** 759-763
[4] Bent J.F,Richards R.W., Gough T.D. Rev Sc Inst (2003) **74,** 4052-4757
[5] J. Bent, et al. Science (2003) **301**, 1691-1695 ; Clarke et al Macromolecules (2006) **39,** 7607-7616 ; McLeish T.C.B. Soft Matter, (2009) **5,** 4426–4432
[6] Qazi S.J.S., R. Rennie A.R., Tucker I., Penfold J., Grillo I (2010) J. Phys. Chem. B, submitted
[7]Egelhaaf S.U., Schurtenberger P. PRL (1999) **82**,2804-2807
[8] Grillo I. COCIS (2009) **14,** 402–408
[9] Impéror-Clerc M., Grillo I. ,Khodakov A.Y., Durand D. , Zholobenko V.L. Chem Commun (2007) 834–836
[10] Perino-Gallice L., Fragneto G., Mennicke U., Salditt T., Rieutord F. Eur. Phys. J. E (2002) **8**, 275–282
[11] E. Del Favero, A. Raudino, P. Brocca, S. Motta, G. Fragneto, M. Corti and L. Cantú Langmuir (2009) **25,** 4190–4197
[12] Cavalcanti L.P., Haas H., Bordallo H.N., Konovalov O., Gutberlet T., Fragneto G. Eur. Phys. J. special Topics, (2007) **141**, 217-221
[13] Ding L., Weiss T. M., Fragneto G., Liu W., Yang L., Huang H. W. Langmuir (2005) **21**, 203-210
[14] Schneck E., Oliveira R.G., Rehfeldt F., Demé B., Brandenburg K., Seydel U., Tanaka M. Physical Review E (2009) **80,** 41929
[15] Appavou M.S., Gibrat G., Bellissent-Funel M.-C., Plazanet M. ,Pieper J. , Buchsteiner A., Annighöfer B. J. Phys.: Condens. Matter (2005) **17**, S3093–S3099
[16] Kohlbrecher J., Bollhalder A., R. Vavrina, Meier G. R. Rev. Sci. Instrum. (2007) **78**, 125101 :1-6

[17] Plazanet M., R. Schweins R., P. Lindner P., Trommsdorff H.P. J. Phys. IV France (2005) **130**, 81–84
[18] Pépy G. and P. Baroni P., J. Appl. Cryst. (2003) **36**, 814-815
[19] Ribaut T., Oberdisse J., Annighofer B., Stoychev I., Fournel B., Sarradee S., and Lacroix-Desmazes P Soft Matter (2009) **5**, 4962–4970
[20] Trickett K., Xing D.Z., Eastoe J., Enick R., Mohamed A., Hollamby M.J., Cummings S., Rogers S.E., Heenan R.K. Langmuir (2010), **26**, 4732–4737
[21] Xu F; Diat O., Gebel G. Morin A., Journal of the Electrochemical Society (2007) **154**, B1389-B1398. Gebel G., Diat O., Escribano S., Mosdale R. Journal of Power Sources (2008) **179**, 132-139.
[22] Keiderling U., Applied Physics A- Materials sciences and processing (2002) **74** S1455-S1457
[23] http://www.ill.eu/instruments-support/instruments-groups/groups/lss/grasp/home/
[24] http://www.ill.eu/instruments-support/computing-for-science/cs-software/all-software/lamp/the-lamp-book/
[25] http://didier.lairez.fr/dokuwiki/doku.php?id=pasinet
[26] http://www.nexusformat.org/Main_Page
[27] Glatter O. Acta Phys. Austriaca 1977a, 47, 83-102; Glatter O. J. Appl. Cryst. (1977) **10**, 415-421
[28] Brunner-popela J., Glatter O. J App Cryst (1997) **30**, 431-442
[29] Svergun D.I. Acta Cryst. A (1994) **50**, 391-402
[30] Svergun D.I., Barberato C. Koch M.H.J. J. Appl. Cryst. (1995) **28**, 768-773
[31] Svergun D.I. Biophys. J. (1999) **76**, 2879-2886
[32] http://www.embl-hamburg.de/ExternalInfo/Research/Sax/software.html
[33] Pedersen J.S., Gerstenberg MC Macromolecules (1996), **29**, 1363-1365
[34] Perdesen J.S. Chapter 15 in Neutrons, X-rays and light: scattering methods applied to soft condensed matter 2002, Ed P. Lindner and Th. Zemb, Elsevier
[35] Oberdisse J., Hine P., Pyckhout-Hintzen W. (2007) Soft Matter, **3**, 476–485
[36] Testard V., Oberdisse J., Ligoure C. Macromolecules (2008) **41**, 7219-7226
[37] McGreevy R.L., Pusztai L. Mol. Simul. (1988) **1** 359
[38] McGreevy R.L., J. Phys.: Condens. Matter (2001) **13**,R877–R913
[39] Kline S.R. J Appl. Cryst. (2006) **39**, 895
[40] http://danse.chem.utk.edu/
[41] Dewhurst C.D., Measurement Science and Technology (2008) **19**, 034007-1-034007-8
[42] Desert S.,Thevenot V., Oberdisse J., Brûlet A. J. Appl. Cryst. (2007) **40**, s471–s473
[43] Brûlet A., Thevenot V., Lairez D. Lecommandoux S., Agut W., Armes S.P., Duc J., Désert S.J. Appl. Cryst. (2008) **41**, 161–166
[44] R.E. Williams, J.M. Rowe Physica B (2002) **311**, 117–122
[45] Mildner D.F.R., Hammouda B., Kline S.R. J. Appl. Cryst. (2005) **38**, 979–987
[46] Koizumi S., Iwase H., Suzuki J.-I., Oku T., Motokawa R, Sasao H., Tanaka H., Yamaguchi D., Shimizu H.M., Takeji Hashimoto T. J. Appl. Cryst. (2007) **40**, s474–s479
[47] Littrell K.C. Nuclear Instruments and Methods in Physics Research A (2004) **529**, 22–27